\documentclass[aps,12pt,superscriptaddress,amsfonts,amssymb,amsmath]{revtex4}
\pdfoutput=1

\usepackage{graphicx}
\usepackage{epsfig}
\usepackage{makeidx}
\usepackage{epstopdf}
\usepackage{natbib}
\usepackage{xcolor}
\usepackage{subcaption}
\usepackage{amsmath}
\usepackage{appendix}

\begin{document}

\title{Generalized local charge conservation in many-body quantum mechanics
}
\author{F.\ Minotti \footnote{Email address: minotti@df.uba.ar}}
\affiliation{Universidad de Buenos Aires, Facultad de Ciencias Exactas y Naturales, Departamento de F\'{\i}sica, Buenos Aires, Argentina}
\affiliation{CONICET-Universidad de Buenos Aires, Instituto de F\'{\i}sica Interdisciplinaria y Aplicada (INFINA), Buenos Aires, Argentina}

\author{G.\ Modanese \footnote{Email address: giovanni.modanese@unibz.it}}
\affiliation{Free University of Bozen-Bolzano \\ Faculty of Engineering \\ I-39100 Bolzano, Italy}

\linespread{0.9}

\begin{abstract}
 
In the framework of the quantum theory of many-particle systems, we study the compatibility of approximated Non-Equilibrium Green Functions (NEGFs) and of approximated solutions of the Dyson equation with a modified continuity equation of the form $\partial_t \langle \rho \rangle+(1-\gamma)\nabla\cdot \langle\mathbf{J} \rangle=0$. A continuity equation of this kind allows the e.m.\ coupling of the system in the extended Aharonov-Bohm electrodynamics, but not in Maxwell electrodynamics. Focusing on the case of molecular junctions simulated numerically with the Density Functional Theory (DFT), we further discuss the re-definition of local current density proposed by Wang et al., which also turns out to be compatible with the extended Aharonov-Bohm electrodynamics.
 
\end{abstract}

\maketitle

\section{Introduction}
It is known from elementary quantum mechanics that the solutions of the  Schr\"odinger equation satisfy a continuity equation for the probability density $\rho$ and the current density $\mathbf{J}=(-ie\hbar/2m)(\psi^*\nabla\psi-\psi\nabla\psi^*)$. Therefore, when the Schr\"odinger equation is applied to a charged particle, we can say that in the ensuing dynamics charge is locally conserved, at least in a probabilistic sense.

When we turn to a many-body theory of condensed matter, if we want to check the validity of the continuity equation we need a formalism able to describe a system with a large number of degrees of freedom -- or actually infinite degrees of freedom if we use a quantum field theory which takes into account all possible virtual processes.

Historically, local conservation laws in many-particle systems have been first systematically analized in the Non-Equilibrium Green Functions (NEGF) formalism by Baym and Kadanoff \cite{baym1961conservation,baym1962self}. They wrote an exact continuity equation (valid at all perturbative orders) involving the one-particle Green function $G$ and the two-particle Green function $G_2$. An alternative approach uses the Dyson equation, which involves $G$ and the self-energy function $\Sigma$ (see e.g.\ \cite{haug2008quantum}). 

We will use both techniques in this work, applying them to fermionic condensed-matter systems,
for which we take into consideration that the evaluation of the NEGFs
requires the introduction of an ultraviolet (UV) cutoff to regularize
integrals in Feynman graphs. It is not proved at present, at least for 
many-particle fermionic systems in three spatial dimensions, that the UV
cutoff can be removed. For this reason, we consider the possibility that the
NEGFs themselves could be strongly dependent on that cutoff. In particular,
the proof of local conservation of charge requires the identification of a
NEGF in which one of its time arguments is evaluated at $t\pm \varepsilon $,
with the NEGF in which the same time argument is evaluated at $t$ ($t$ is a
generic time, and the symbol $\varepsilon $ represents an infinitesimal
magnitude). However, if the NEGF happens to be cutoff dependent, its time
dependence can be determined only for time intervals larger than those
imposed by the cutoff, so that $\varepsilon $ cannot be arbitrarily small.
 
In this way, introducing also some reasonable approximations and physical cutoffs, we obtain generalized local conservation laws of the form
\begin{equation}
    \partial_t \langle \rho \rangle+(1-\gamma)\nabla\cdot \langle\mathbf{J} \rangle=0
\label{mod_cons}
\end{equation}
that we call ``$\gamma$-models'', in which the positive  non dimensional parameter $\gamma$ is normally very small, $\gamma \ll 1$. 

These modified conservation laws are not compatible with the Maxwell equations, but are fully compatible with the extended electrodynamics of Aharonov-Bohm \cite{ohmura1956new,aharonov1963further,van2001generalisation,hively2012toward,modanese2017electromagnetic,hively2019classical,minotti2021quantum,minotti2023aharonov}. We gave previous examples of simple $\gamma$-models in \cite{EPJC2023}. Actually, in A.-B.\ electrodynamics it is also possible to handle cases where there are major modifications of the local conservation equation, like for the Schr\"odinger equation in the presence of non-local potentials or the fractional Schr\"odinger equation \cite{modanese2018time,lenzi2024}.

It is interesting to recall the motivations of the work by Baym and Kadanoff and compare them with our present approach. Baym and Kadanoff thought it was very important to build all conservation laws into the structure of the approximations used to compute the Green functions. They thus devised a general method for generating ``conserving approximations'', obtained by replacing $G_2$ by suitable functionals of $G$.

In our present work we only consider a few specific examples of conserving approximations, but referred to the more general conservation law \eqref{mod_cons}. For this purpose, we use the NEGF formalism in Sect.\ \ref{strong-argument} and the Dyson equation in Sect.\ \ref{Alternative-approach}. 

In Sect.\ \ref{cabra-wang} we recall for comparison the results obtained by Cabra et al.\ (\cite{cabra2018simulation} and refs.). Cabra et al.\ write a continuity equation derived from the Dyson equation and including a source term, namely
\begin{equation}
    \partial_t \rho(\mathbf{r},t)+ \nabla\cdot \mathbf{J} (\mathbf{r},t)=P(\mathbf{r},t)
\label{contin_cabra}
\end{equation}
with
\begin{equation}
    P(\mathbf{r},t)=2 \text{Re}\int d\mathbf{r}_1 \int dt_1 \left[ G^<(\mathbf{r},t;\mathbf{r}_1,t_1) \Sigma^a(\mathbf{r}_1,t_1;\mathbf{r},t)+  G^r(\mathbf{r},t;\mathbf{r}_1,t_1) \Sigma^<(\mathbf{r}_1,t_1;\mathbf{r},t) \right]
\label{P_cabra}
\end{equation}
where the upper indices of the Green function $G$ and the self-energy $\Sigma$ denote the boundary conditions that they satisfy.

They discuss the form and magnitude of the source term which result from their numerical DFT (Density Functional Theory) approximation, with special reference to molecular junctions. In spite of some technical limitations, the effectiveness of DFT can hardly be questioned, given its accurate results for the energy spectra, dipole momenta and other physical quantities. A generalized local conservation law (not necessarily in the form of a $\gamma$-model) therefore seems to be needed also in their approach.

Additionally, in Sect.\ \ref{cabra-wang} we briefly consider the criticism expressed by Cabra et al.\ in \cite{cabra2018simulation} towards the method by Li, Wang et al.\ \cite{li2008definition,zhang2011first}; our conclusion is that both approaches can be reconciled in the framework of the extended A.-B.\  electrodynamics.

\section{On local conservation of charge in the NEGF formalism}
\label{strong-argument}

As mentioned in the Introduction, there is a strong argument for local charge conservation in many-body
quantum mechanics based on the formalism by Baym and Kadanoff of non-equilibrium Green functions (NEGFs).
The argument is based on Eq.\ (21) in the paper \cite{baym1961conservation}, which we rewrite in a
slightly different form, as derived from Eqs.\ (3) in \cite{baym1962self}, using the same notation to facilitate comparison with that reference:
\begin{eqnarray}
&&\left[ \frac{\partial }{\partial t_{1}}+\frac{\partial }{\partial
t_{1^{\prime }}}+\frac{1}{2im}\left( \nabla _{1}+\nabla _{1^{\prime
}}\right) \cdot \left( \nabla _{1}-\nabla _{1^{\prime }}\right) \right]
iG\left( 1,1^{\prime }\right)  \notag \\
&&+\int \left[ G\left( 1,\overline{1}\right) U\left( \overline{1},1^{\prime
}\right) -U\left( 1,\overline{1}\right) G\left( \overline{1},1^{\prime
}\right) \right]  \notag \\
&=&\pm i\int \left[ V\left( 1-\overline{1}\right) G_{2}\left( 1\overline{1}
,1^{\prime }\overline{1}_{+}\right) -G_{2}\left( 1\overline{1}_{-},1^{\prime
}\overline{1}\right) V\left( \overline{1}-1^{\prime }\right) \right] ,
\label{BK}
\end{eqnarray}
where the upper sign applies to bosons and the lower sign to fermions. In this expression the arguments $1$, $1'$, $\bar{1}$ represent in compact notation the sets of coordinates $(t_1,\mathbf{x}_1)$, $(t_{1'},\mathbf{x}_{1'})$,$(t_{\bar{1}},\mathbf{x}_{\bar{1}})$, respectively, and the four-integral is understood to be performed on the variables with a macron, like $\bar{1}$ in this case. Furthermore, $\overline{1}_{+}$ means $
\mathbf{x}_{\overline{1}}$, $t_{\overline{1}_{+}}$, with $t_{\overline{1}_{+}}$ infinitesimally larger than $t_{\overline{1}}$, and correspondingly, $t_{\overline{1}_{-}}$ infinitesimally smaller than $t_{\overline{1}}$.

It was shown in \cite{baym1961conservation} that
\begin{eqnarray*}
&&\underset{1^{\prime }\rightarrow 1_{+}}{\lim }\left[ \frac{\partial }{
\partial t_{1}}+\frac{\partial }{\partial t_{1^{\prime }}}+\frac{1}{2im}
\left( \nabla _{1}+\nabla _{1^{\prime }}\right) \cdot \left( \nabla
_{1}-\nabla _{1^{\prime }}\right) \right] iG\left( 1,1^{\prime }\right) \\
&=&\pm \left[ \frac{\partial \left\langle n\left( 1\right) \right\rangle }{
\partial t}+\nabla \cdot \left\langle \mathbf{j}\left( 1\right)
\right\rangle \right] ,
\end{eqnarray*}
where the averages are taken on a grand-canonical ensemble. It also was shown in \cite{baym1961conservation} that in the same limit the integral $\int \left[ G\left( 1,\overline{1}\right) U\left( \overline{1},1^{\prime
}\right) -U\left( 1,\overline{1}\right) G\left( \overline{1},1^{\prime
}\right) \right]$
goes to zero.

The limit of the r.h.s.\ of eq.\ \eqref{BK} when $1^{\prime }\rightarrow 1_{+}$ is 
\begin{equation}
   \pm i\int V\left( 1-\overline{1}\right) \left[ G_{2}\left( 1\overline{1},1
\overline{1}_{+}\right) -G_{2}\left( 1\overline{1}_{-},1\overline{1}\right) 
\right]  ,
\end{equation}
because the potential $V$ is instantaneous: $V\left( 1-\overline{1}\right)=V\left(\mathbf{
x}_{1}-\mathbf{x}_{\overline{1}}\right)\delta(t_{1}-t_{\overline{1}})$.

Recalling that $\overline{1}$ symbolizes the space time coordinates $\mathbf{
x}_{\overline{1}}$, $t_{\overline{1}}$, that $\overline{1}_{\pm }$ means $
\mathbf{x}_{\overline{1}}$, $t_{\overline{1}_{\pm} }$, and that in the Baym and Kadanoff formalism the time integration is along the imaginary time $0<it<\beta $, we can write 
\begin{eqnarray*}
G_{2}\left( 1\overline{1},1\overline{1}_{+}\right) -G_{2}\left( 1\overline{1}
_{-},1\overline{1}\right) &=&\left[ \tau \frac{\partial }{\partial t_{2}}
G_{2}\left( 1\overline{1},12\right) -\left( -\tau \right) \frac{\partial }{
\partial t_{2}}G_{2}\left( 12,1\overline{1}\right) \right] _{2=\overline{1}}
\\
&=&\tau \frac{\partial }{\partial t_{\overline{1}}}G_{2}\left( 1\overline{1}
,1\overline{1}\right) ,
\end{eqnarray*}
where $\tau $ is of infinitesimal magnitude with units of imaginary time,
used to express $t_{\overline{1}_{\pm}}=t_{\overline{1}} \pm\tau $.

We thus have, returning also to standard units, ($\hbar \neq 1$),
\begin{equation}
\frac{\partial \left\langle n\left( 1\right) \right\rangle }{\partial t}
+\nabla \cdot \left\langle \mathbf{j}\left( 1\right) \right\rangle =\frac{
i\tau }{\hbar }\int V\left( 1-\overline{1}\right) \frac{\partial }{\partial
t_{\overline{1}}}G_{2}\left( 1\overline{1},1\overline{1}\right) .
\label{LNC}
\end{equation}

In order to consider if $\tau $ could be taken as arbitrarily small, we recall that the NEGF formalism includes an UV cutoff to regularize integrals over momentum and energy in Feynman graphs, and that in 3-D it is not proven that the cutoff can be removed \cite{salmhofer2019renormalization}. Consequently, the high-frequency (short time) behavior of NEGFs could be strongly sensitive to that cutoff, and could not be extrapolated to arbitrarily short time intervals. For a fermionic condensed matter system such as a conductor or semiconductor, a natural cutoff should allow the inclusion of pertinent
processes like, for instance, multiple electron-hole production and destruction in internal lines of Feynman graphs.

In order to make some rough estimations, and to show that even a rather small $\tau$ can have observable effects, we consider a rather extreme cutoff that
excludes virtual electron-positron creation processes in internal lines, so
we can assume that $\left\vert \tau \right\vert $ cannot be smaller than $
\hbar /E_{V}$, with $E_{V}\sim 1$ MeV.

With these considerations, we have from (\ref{LNC})
\begin{eqnarray*}
\frac{\partial \left\langle n\left( 1\right) \right\rangle }{\partial t}
+\nabla \cdot \left\langle \mathbf{j}\left( 1\right) \right\rangle &\sim &-
\frac{1}{E_{V}}\int V\left( 1-\overline{1}\right) \frac{\partial }{\partial
t_{\overline{1}}}G_{2}\left( 1\overline{1},1\overline{1}\right) \\
&=&-\frac{1}{E_{V}}\int V\left( \mathbf{x}_{1}-\mathbf{x}_{\overline{1}
}\right) \left[ \frac{\partial }{\partial t_{\overline{1}}}G_{2}\left( 1
\overline{1},1\overline{1}\right) \right] _{t_{\overline{1}}=t_{1}}d^{3}x_{\overline{1}}.
\end{eqnarray*}

We can further write
\begin{equation*}
\frac{\partial }{\partial t_{\overline{1}}}G_{2}\left( 1\overline{1},1
\overline{1}\right)= \left\langle n\left( 1\right) \frac{\partial
n\left( \overline{1}\right) }{\partial t_{\overline{1}}}\right\rangle,
\end{equation*}
to have
\begin{eqnarray*}
\frac{\partial \left\langle n\left( 1\right) \right\rangle }{\partial t}
+\nabla \cdot \left\langle \mathbf{j}\left( 1\right) \right\rangle &\sim &-
\frac{1}{E_{V}}\int V\left( \mathbf{x}_{1}-\mathbf{x}_{\overline{1}}\right)
\left\langle n\left( \mathbf{x}_{1},t_{1}\right) \frac{\partial n\left( 
\mathbf{x}_{\overline{1}},t_{1}\right) }{\partial t_{1}}\right\rangle
d^{3}x_{\overline{1}} \\
&\sim &\frac{1}{E_{V}}\frac{e^{2}}{4\pi \varepsilon _{0}r_{c}}\left\langle
n\left( 1\right) \right\rangle \frac{\partial \left\langle n\left( 1\right)
\right\rangle }{\partial t_{1}}\frac{4\pi }{3}r_{c}^{3},
\end{eqnarray*}
where $r_{c}$ denotes a correlation distance.

Since we have with good approximation 
\begin{equation*}
\frac{\partial \left\langle n\left( 1\right) \right\rangle }{\partial t_{1}}
\simeq -\nabla \cdot \left\langle \mathbf{j}\left( 1\right) \right\rangle ,
\end{equation*}
we finally obtain a $\gamma $ model: 
\begin{equation*}
\frac{\partial \left\langle n\left( 1\right) \right\rangle }{\partial t}
+\nabla \cdot \left\langle \mathbf{j}\left( 1\right) \right\rangle =\gamma
\nabla \cdot \left\langle \mathbf{j}\left( 1\right) \right\rangle ,
\end{equation*}
with 
\begin{equation}
\gamma \sim -\frac{1}{E_{V}}\frac{e^{2}}{4\pi \varepsilon _{0}r_{c}}
\left\langle n\right\rangle \frac{4\pi }{3}r_{c}^{3}.  \label{g_Baym}
\end{equation}

For the conducting electrons in a metal we have that the correlation given
by the screened Coulomb interaction has \cite{ashcroft_mermin1976} 
\begin{equation*}
r_{c}\simeq \left[ \frac{3}{4\pi \left\langle n\right\rangle }\right] ^{1/3},
\end{equation*}
so that with $E_{V}\simeq 1$ MeV, for Cu, with $\left\langle n\right\rangle
\simeq 8.5\times 10^{28}$ m$^{-3}$, we have $\gamma \sim -10^{-5}$. Since
the number density of conducting electrons in most metals is of the same
order, we expect similar values of $\gamma $ for them.

The functions used and their equations and boundary conditions are defined
only on the imaginary time domain $0<it<\beta $. Physical magnitudes are
obtained after analytic continuation of the calculated expressions. This is
usually done by first considering the imaginary time domain to be $0<i\left(t-t_{0}\right) <\beta $, with $t_{0}$ real, and after the analytic continuation is obtained, the limit $t_{0}\rightarrow -\infty $ taken \cite{kadanoff_baym1962,haug2008quantum}.

This method leads in general to the conclusion that the NEGFs defined in the imaginary time tend when $t_{0}\rightarrow -\infty $ to the physical NEGFs defined for real times. For this reason we can consider the relations
obtained above as valid also for the physical magnitudes.

\section{Alternative approach}
\label{Alternative-approach}

In the applications the equation for the one-particle NEGF $G\left(
1,2\right) $\ is usually written in terms of the self-energy $\Sigma \left(
1,2\right) $, instead of the two-particle NEGF $G_{2}\left( 12,34\right) $
(Dyson equation) \cite{baym1962self}:
\begin{equation}
\left( i\frac{\partial }{\partial t_{1}}+\frac{\nabla _{1}^{2}}{2m}\right)
G\left( 1,2\right) =\delta \left( 1-2\right) +\int U\left( 1,\overline{1}
\right) G\left( \overline{1},2\right) +\int \Sigma \left( 1,\overline{1}
\right) G\left( \overline{1},2\right) .  \label{Dyson}
\end{equation}

From this equation and its adjoint, we can obtain in a similar manner as above the equation
\begin{equation}
\frac{\partial \left\langle n\left( 1\right) \right\rangle }{\partial t}
+\nabla \cdot \left\langle \mathbf{j}\left( 1\right) \right\rangle =\pm 
\underset{2\rightarrow 1_{+}}{\lim }\int \left[ \Sigma \left( 1,\overline{1}
\right) G\left( \overline{1},2\right) -G\left( 1,\overline{1}\right) \Sigma
\left( \overline{1},2\right) \right] .  \label{con_Sigma}
\end{equation}

It is important to mention that in \cite{baym1962self} it is proved that if 
\begin{equation*}
    \Sigma(1,2)=\frac{\delta \Phi\left[G\right]}{\delta G\left(2,1_{+}\right)},
\end{equation*}
where $ \Phi\left[G\right]$ is a gauge-invariant functional of $G$, then 
\begin{equation*}
   \int \left[ \Sigma \left( 1,\overline{1}
\right) G\left( \overline{1},1\right) -G\left( 1,\overline{1}\right) \Sigma
\left( \overline{1},1\right) \right]=0.  
\end{equation*}
However, this expression differs from the right-hand side of Eq. (\ref{con_Sigma}), in which $t_{2}$ is evaluated at $t_{1_{+}}$ and not at $t_{1}$, so that an argument similar to that used in the previous section indicates that the right-hand side of Eq. (\ref{con_Sigma}) need not be necessarily zero.

To proceed, and to work with magnitudes defined in real time, the time part of the
integrals in the right-hand sides of these equations is taken on the Keldysh
contour \cite{haug2008quantum}, so that a more explicit expression of the right-hand side of (\ref
{con_Sigma}) is
\begin{eqnarray*}
&&\pm \int_{-\infty }^{\infty }dt_{\overline{1}}\int d^{3}x_{\overline{1}}
\left[ \Sigma ^{R}\left( 1,\overline{1}\right) G^{<}\left( \overline{1}
,1\right) +\Sigma ^{<}\left( 1,\overline{1}\right) G^{A}\left( \overline{1}
,1\right) \right. \\
&&\;\;\;\;\;\;\;\;\;\;\;\;\left. -G^{R}\left( 1,\overline{1}\right) \Sigma
^{<}\left( \overline{1},1\right) -G^{<}\left( 1,\overline{1}\right) \Sigma
^{A}\left( \overline{1},1\right) \right] ,
\label{more-expl}
\end{eqnarray*}
where retarded (R), advanced (A) and lesser ($<$) functions are explicitly indicated.

In order to make a rough estimation we consider the Hartree-Fock approximation in the static limit for the self-energies, in which $\Sigma
^{\lessgtr }$ are neglected, and 
\begin{equation*}
\Sigma ^{R}\left( 1,\overline{1}\right) =\Sigma ^{A}\left( 1,\overline{1}
\right) =\Sigma _{HF}\left( \mathbf{x}_{1}-\mathbf{x}_{\overline{1}}\right)
\delta \left( t_{1}-t_{\overline{1}}\right) ,
\end{equation*}
so that 
\begin{equation*}
\frac{\partial \left\langle n\left( 1\right) \right\rangle }{\partial t}
+\nabla \cdot \left\langle \mathbf{j}\left( 1\right) \right\rangle \simeq
\int d^{3}x_{\overline{1}}\Sigma _{HF}\left( \mathbf{x}_{1}-\mathbf{x}_{
\overline{1}}\right) \left[ G^{<}\left( \overline{1},1\right) -G^{<}\left( 1,
\overline{1}\right) \right] _{t_{\overline{1}}=t_{1}}.
\end{equation*}

In terms of the second-quantized, Heisenberg representation, particle
creation and annihilation operators $\psi ^{\dagger }\left( \mathbf{x}
,t\right) $ and $\psi \left( \mathbf{x},t\right) $ we have
\begin{equation*}
G^{<}\left( \overline{1},1\right) =i\left\langle \psi ^{\dagger }\left( 
\mathbf{x}_{1},t_{1}\right) \psi \left( \mathbf{x}_{\overline{1}},t_{
\overline{1}}\right) \right\rangle .
\end{equation*}

Thus, if we further consider a rapid decay of $\Sigma _{HF}$ with $
\left\vert \mathbf{x}_{1}-\mathbf{x}_{\overline{1}}\right\vert $, we can
Taylor expand the particle operators around $\mathbf{x}_{1}$ to approximate
(summation over repeated indices is assumed)
\begin{eqnarray*}
\left[ G^{<}\left( \overline{1},1\right) -G^{<}\left( 1,\overline{1}\right) 
\right] _{t_{\overline{1}}=t_{1}} &\simeq &i\mathbf{r}\cdot \left[
\left\langle \psi ^{\dagger }\nabla \psi \right\rangle -\left\langle \nabla
\psi ^{\dagger }\psi \right\rangle \right] \\
&&+\frac{i}{2}r_{l}r_{m}\left[ \left\langle \psi ^{\dagger }\frac{\partial
^{2}\psi }{\partial x_{1l}\partial x_{1m}}\right\rangle -\left\langle \frac{
\partial ^{2}\psi ^{\dagger }}{\partial x_{1l}\partial x_{1m}}\psi
\right\rangle \right] ,
\end{eqnarray*}
where $\mathbf{r}=\mathbf{x}_{\overline{1}}-\mathbf{x}_{1}$, and all
particle operators are functions of $\left( \mathbf{x}_{1},t_{1}\right) $.

For $\Sigma _{HF}\left( \mathbf{x}_{1}-\mathbf{x}_{\overline{1}}\right)
=\Sigma _{HF}\left( \left\vert \mathbf{x}_{1}-\mathbf{x}_{\overline{1}
}\right\vert \right) $ integration over the angular dependence of $\mathbf{r}
$ yields (in standard units)
\begin{eqnarray}
\frac{\partial \left\langle n\left( 1\right) \right\rangle }{\partial t}
+\nabla \cdot \left\langle \mathbf{j}\left( 1\right) \right\rangle &\simeq
&\pm \frac{2\pi i}{3\hbar }\left[ \left\langle \psi ^{\dagger }\nabla
^{2}\psi \right\rangle -\left\langle \nabla ^{2}\psi ^{\dagger }\psi
\right\rangle \right] \int_{0}^{\infty }\Sigma _{HF}\left( r\right) r^{4}dr 
\notag \\
&=&\pm \frac{2\pi i}{3\hbar }\nabla \cdot \left[ \left\langle \psi ^{\dagger
}\nabla \psi \right\rangle -\left\langle \nabla \psi ^{\dagger }\psi
\right\rangle \right] \int_{0}^{\infty }\Sigma _{HF}\left( r\right) r^{4}dr 
\notag \\
&=&\mp \frac{4\pi m}{3\hbar ^{2}}\nabla \cdot \left\langle \mathbf{j}\left(
1\right) \right\rangle \int_{0}^{\infty }\Sigma _{HF}\left( r\right) r^{4}dr.
\label{g-Sigma}
\end{eqnarray}

We thus obtain also in this approximation a $\gamma $ model with
\begin{equation*}
\gamma \simeq \mp \frac{4\pi m}{3\hbar ^{2}}\int_{0}^{\infty }\Sigma
_{HF}\left( r\right) r^{4}dr.
\end{equation*}

As a simple example, for a Fermi gas with screened Coulomb interactions
we can approximate \cite{datta1995}
\begin{equation*}
\Sigma _{HF}\left( r\right) =-\frac{e^{2}}{8\pi ^{3}\varepsilon _{0}r^{4}}
\exp \left( -k_{F}r\right) \left[ \sin \left( k_{F}r\right) -k_{F}r\cos
\left( k_{F}r\right) \right] ,
\end{equation*}
where $k_{F}=\left( 3\pi ^{2}\left\langle n\right\rangle \right) ^{1/3}=
\sqrt{2mE_{F}}/\hbar $ is the Fermi wavelength corresponding to the Fermi
energy $E_{F}$. We thus have
\begin{eqnarray*}
\gamma  &\simeq &-\frac{me^{2}}{12\pi ^{2}\varepsilon _{0}\hbar ^{2}k_{F}}=-
\frac{e^{2}k_{F}}{24\pi ^{2}\varepsilon _{0}E_{F}} \\
&=&-\frac{me^{2}}{4\left( 3\pi ^{2}\right) ^{4/3}\varepsilon _{0}\hbar
^{2}\left\langle n\right\rangle ^{1/3}}\simeq -\left( \frac{2.7\times 10^{26}
\text{m}^{-3}}{\left\langle n\right\rangle }\right) ^{1/3}.
\end{eqnarray*}
For Cu this gives $\gamma \simeq -0.15$, which is only indicative due to the
approximations made. Using the formalism in the previous section, a  $\gamma$ of similar magnitude could be obtained with an UV cutoff of approximately 100 eV.

In any case, it is interesting to note that in the first-principle calculations
by Zhang et al. \cite{zhang2011first} their Fig. 4 shows the
conventional and non-local currents. The divergence of the non-local current
is minus the divergence of the conventional current, times a factor of the
order of 0.15, thus satisfying a $\gamma $ model in which $\gamma $ has the
correct sign and similar magnitude.

\section{Generalized continuity equations in the approaches by Cabra et al.\ and Wang et al.}
\label{cabra-wang}

In \cite{cabra2018simulation} Cabra et al.\ give a short review of techniques and results in the theoretical characterization of local properties of molecular junctions (as opposed to properties which characterize the response of the junction as a whole). Their focus is on the calculation and simulation of local currents, also with the aim of clarifying possible misconceptions about these physical quantities. They give the expression of local currents in terms of NEGFs (both in orbital space and real space) and discuss local conservation conditions.

The modified continuity equation obtained, which includes a source term $P(\mathbf{r},t)$, has been already reported in our present Introduction (eqs.\ \eqref{contin_cabra}, \eqref{P_cabra}).
Note that the source term $P(\mathbf{r},t)$ is precisely the r.h.s.\ of our eq.\ \eqref{con_Sigma}, in its explicit version \eqref{more-expl} obtained using the Keldish method.

In steady-state conditions one can obtain a modified integral conservation law for the total charge in an arbitrary volume $V$, in the form
\begin{equation}
    \oint_{\partial V} J_k(\mathbf{r})d\sigma_k=\int_V P(\mathbf{r})d^3r
\end{equation}
This shows, in their own words, that ``local currents within the junction are not conserved because of electron density production induced by the source term $P$.''  When $V$ is extended to the whole junction, global charge conservation is recovered, thanks to the fact that the integral of $P$ over all space is zero.

Cabra et al.\ also notice that the local unbalance of charge depends on the partition defined between the system (the junction) and the external contacts. This issue has been already discussed in \cite{walz2015local} and is related in our opinion to a partial failure in local gauge invariance when the interaction of a dynamical system with its environment is  added to an intrinsically symmetric Lagrangian (see Sect.\ \ref{conc}).

Cabra et al.\ object to the idea of imposing local current conservation as proposed in \cite{li2008definition,zhang2011first} because of the ambiguities involved in representing the source term $P$ as the divergence of a local flux.

It is worth recalling briefly here the argument given by Wang et al.\ in \cite{li2008definition,zhang2011first}. They write the Schr\"odinger equation as
\begin{equation}
    i\hbar\partial_t \psi(\mathbf{r},t)= -\frac{\hbar^2}{2m} \nabla^2 \psi(\mathbf{r},t) + \int d^3r' V(\mathbf{r},\mathbf{r}') \psi(\mathbf{r}',t)
\end{equation}
Using the conjugate equation they obtain for the conventional charge density $\rho$ and current density $\mathbf{J}$ the modified continuity equation
\begin{equation}
    -\partial_t \rho(\mathbf{r},t)=\nabla\cdot \mathbf{J}(\mathbf{r},t)+\rho_n(\mathbf{r},t)
\end{equation}
where
\begin{equation}
    \rho_n(\mathbf{r},t)=\frac{e}{i\hbar} \int d^3r' [\psi(\mathbf{r},t)V^*(\mathbf{r},\mathbf{r}') \psi(\mathbf{r}',t)-\text{c.c.}]
\end{equation}

If the potential is local, the quantity $\rho_n$ vanishes; if not, the usual continuity condition is not satisfied. (A typical example is given by the exchange-correlation energy functional beyond the local density approximation.) Considering for simplicity the steady state without time dependence, Wang et al.\ then propose to define a new locally conserved current density with zero divergence:
\begin{equation}
    \mathbf{J}_{cons}=\mathbf{J}+\mathbf{J}_n
\end{equation}
where
\begin{equation}
    \mathbf{J}_n(\mathbf{r})=-\nabla \varphi_n(\mathbf{r})
\end{equation}
and $\varphi_n$ satisfies the Poisson equation
\begin{equation}
    \nabla^2 \varphi_n(\mathbf{r})=-\rho_n(\mathbf{r})
\end{equation}
(Note that $\mathbf{J}_{cons}$ is simply called $\mathbf{J}$ in \cite{li2008definition}.)

In order to support their definition, Wang et al.\ prove that the locally conserved current $\mathbf{J}_{cons}$ is the same that would be obtained from the Landauer-B\"uttiker formula of quantum transport.

While Cabra et al.\ and Wang et al.\ agree on the fact that the source term in the modified continuity equation (which they call respectively $P$ and $\rho_n$) cannot be disregarded, 
the definition of locally-conserved current given by Wang et al.\ is criticized for two reasons:

(1) \emph{It would imply an unrealistic (especially in anisotropic media) fixed proportionality between} $\mathbf{J}_{cons}$ \emph{and the local electric field.} However, this proportionality is not explicitly stated in \cite{li2008definition}, \cite{zhang2011first}. It might have been inferred in \cite{cabra2018simulation} as a consequence of the Maxwell equations, but this inference is not correct in our opinion, in view of the extended electrodynamics with Aharonov-Bohm equations, see below.

(2) \emph{The condition} $\nabla\cdot \mathbf{J}_{cons}=0$ \emph{cannot define} $\mathbf{J}_{cons}$ \emph{in a unique way because it leaves the freedom to add an arbitrary curl to} $\mathbf{J}_{cons}$.  Again, the extended e.m.\ field equations can resolve this problem.

We next explain in more detail our replies to the objections above.

(1) The first equation of the extended Aharonov-Bohm electrodynamics is
\begin{equation}
    \nabla \cdot \mathbf{E}=\frac{\rho}{\varepsilon_0}-\partial_t S
\end{equation}
where $S$ is the scalar field generated by the ``extra-current'' $I=\partial_t\rho+\nabla\cdot\mathbf{J}$ (the same quantity which is called source term by Cabra et al.\ and Wang et al.\, and denoted respectively by $P$ and $\rho_n$), according to the equation $\Box S=\mu_0 I$ ($\Box$ is the D'Alembert operator). In a steady state, the divergence of $\mathbf{E}$ is uniquely determined by the real electric charge density $\rho$ and there is no proportionality relation between $\mathbf{E}$ and $J_{cons}$.

(2) The third equation of the extended A.-B.\ electrodynamics is
\begin{equation}
    \nabla\times \mathbf{B}=\mu_0\varepsilon_0 \partial_t \mathbf{E}+\mu_0 \mathbf{J}+\nabla S
\end{equation}
This shows that the additional current density $\nabla S/\mu_0$ caused by the failure of local conservation indeed has the form proposed by Wang et al. In this connection it is also interesting to note that in the extended A.-B.\ electrodynamics, just like in Maxwell electrodynamics, the magnetic field can be computed from the equation
\begin{equation}
    \Box \mathbf{B}=\mu_0 \nabla \times \mathbf{J}
\end{equation}
and is thus independent from $S$. This is consistent with the fact that the additional current required to restore local conservation is the pure gradient $\nabla S/\mu_0$ and there is no reason to worry about its magnetic effects. On the opposite, adding any arbitrary curl to it would have observable effects on $\mathbf{B}$.

\section{Conclusions}
\label{conc}

In this paper we have investigated the possible occurrence of some modified continuity relations in approximate solutions of the NEGF theory and of the Dyson equation for many-particle quantum systems.

A strong objection to the possibility of local non-conservation is based on
the fundamental nature of quantum gauge field theories, in particular QED.
In QED the condition of local U(1) symmetry leads directly to the e.m.\ field
as the gauge field, satisfying Maxwell equations and thus local conservation.
In fact, however, local conservation of current results, using Noether theorem, directly
from the global U(1) invariance of the action.
In a forthcoming paper, we will discuss this issue in detail for the
more general case of Lagrangians which depend also on the second derivatives
of the fields.

It is hard, in most cases, to escape the objection above, due to the fundamental nature of
Noether theorem.
A known possibility is that the renormalized theory no longer has the
global U(1) invariance of the bare theory. Can the UV cutoff
spoil the action invariance? This question is still open, see our evaluations in the NEGF formalism presented in Sect.\ \ref{strong-argument}.

An alternative possibility is that the field equations derived from a U(1)-invariant Lagrangian
are not strictly satisfied
in some region, so that the same derivation of Noether theorem gives
a continuity equation with a source term.

This could be the case, for instance, if the quantum system undergoes an
evolution not described by the equations of motion, such as the case of the
collapse of the wave function. The consideration that measuring a tunneling
current could be thought of as determining the localization of the electrons
at either side of the barrier, and thus collapsing their wave function, goes
along this possibility, and is a justification for the type of $\gamma$-model derived in \cite{EPJC2023}.

Another underlying mechanism
for a possible failure of local conservation 
is the existence of non-local interactions. Although all interactions considered in the
usual theories are local, non-local interactions can appear in the
effective action of a theory as a result of the exchange interaction. 
An example is the calculation in \cite{lai2019charge}  with an exchange
correlation potential, giving the same result as in our eq.\ \eqref{g_Baym}, apart from a 1/2 factor due to a different definition.
The NEGF formalism does in fact result in equations with effective non-local interactions, like that represented by the self-energy term in the equation for the
one-particle NEGF.

\bibliographystyle{ieeetr}
\bibliography{generalized_conservation}

\end{document}